\documentclass[conference]{IEEEtran}  
\usepackage[dvips]{graphicx}
\usepackage{epsfig,amssymb,amsmath}
\newcommand{\beq}{\begin{equation}}
\newcommand{\eeq}{\end{equation}}
\newcommand{\beqn}{\begin{eqnarray}}
\newcommand{\eeqn}{\end{eqnarray}}
\def\bmath#1{\mbox{\boldmath$#1$}}

\long\def\symbolfootnote[#1]#2{\begingroup%
\def\thefootnote{\fnsymbol{footnote}}\footnote[#1]{#2}\endgroup}

\title{Radio Interferometric Calibration Using The SAGE Algorithm}
\author{\authorblockN{Sarod Yatawatta$^{1,2,3}$, Saleem Zaroubi$^1$, Ger de Bruyn$^{1,2}$, Leon Koopmans${^1}$ and Jan Noordam$^2$}
}
\begin{document}

\maketitle

\symbolfootnote[0]{$^1$Kapteyn Astronomical Institute, University of Groningen, Groningen, The Netherlands}
\symbolfootnote[0]{$^2$ASTRON, Dwingeloo, The Netherlands}
\symbolfootnote[0]{$^3$yatawatta@astron.nl}
\symbolfootnote[0]{Appearing in 13th IEEE DSP Workshop, Marco Island, FL, Jan. 2009.}
\begin{abstract}
Radio Interferometry is an essential method for astronomical observations. Self-calibration techniques have increased the quality of the radio astronomical observations (and hence the science) by orders of magnitude. Recently, there is a drive towards sensor arrays built using inexpensive hardware and distributed over a wide area acting as radio interferometers. Calibration of such arrays poses new problems in terms of computational cost as well as in performance of existing calibration algorithms. We consider the application of the Space Alternating Generalized Expectation Maximization (SAGE) \cite{Fess94} algorithm for calibration of radio interferometric arrays. Application to real data shows that this is an improvement over existing calibration algorithms that are based on direct, deterministic non linear optimization. As presented in this paper, we can improve the computational cost as well as the quality of the calibration using this algorithm.
\end{abstract}

\begin{keywords}
Radio astronomy, Array calibration, EM algorithm, SAGE algorithm
\end{keywords}

\section{Introduction}
Radio synthesis arrays have greatly benefited by self calibration techniques invented during the last 30 years. Calibration refers to estimation of errors introduced by the instrument (and also the propagation path, such as the ionosphere), and correction for such errors,  before any imaging is done. At the beginning of radio astronomy, calibration was done by observing a known celestial object (called the external calibrator), in addition to the part of the sky being observed. This was improved by self-calibration, which is essentially using the observed sky  itself for the calibration. Therefore, self calibration entail considering both the sky as well as the instrument as unknowns. Nevertheless, by iteratively refining the sky and the instrument model, the quality of the calibration was improved by orders of magnitude in comparison to using an external calibrator.

 From a signal processing perspective, calibration is essentially the Maximum Likelihood (ML) estimation of the instrument and sky parameters using a non linear optimization technique such as the Levenberg Marquardt \cite{Lev44},\cite{Mar63} (LM) algorithm. An in depth overview of existing calibration techniques are given in \cite{Boonstra03},\cite{AJV04}. However, with such techniques, we have reached a limit in sensitivity that can be achieved using present radio interferometers. This is because the achievable sensitivity is limited by the receiver collecting area itself. Moreover, with finite computational cost, there is a bound in the performance of existing algorithms. Therefore, there is a drive towards  building large, distributed, sensor arrays (SKA\cite{SKA}, LOFAR \cite{LOFAR}) that increase the collecting area and hence the sensitivity, enabling new scientific research. However, this also increases the parameter space that needs to be calibrated and hence, the required computational cost. A detailed analysis of calibration of LOFAR, which can be considered as a pathfinder for the next generation interferometric arrays, can be found in \cite{Jeffs06}.
 Although in \cite{Jeffs06}, limitations (such as the Cramer Rao bound) of existing calibration techniques are defined, there is little work existing on improving the computational cost or the speed of convergence of such techniques. 

This paper describes the application of the Expectation Maximization \cite{DLR} (EM) algorithm for radio interferometric calibration. The EM algorithm was first presented as a solution to maximum likelihood estimation when the complete data is not observed. Since then, EM type algorithms have been widely used as an iterative method for ML estimation, with either faster convergence or reduced computational cost. The essential property of the EM algorithm is that the likelihood can only increase in each iteration. The SAGE algorithm was presented to improve the plain EM algorithm in terms of speed of convergence and  has been successfully applied in diverse signal processing applications, such as medical imaging \cite{KSR99}, communications systems \cite{Logo00}, etc.
In contrast to \cite{Jeffs06}, in this paper we focus on finding algorithms that improve the computational cost and quality of calibration.  Naturally, in order to improve the performance of the ML estimation, we choose the EM algorithm, and in particular the SAGE algorithm. Therein lies the novelty of this paper.


{\em Notation}: Lower case bold letters refer to column vectors (e.g. ${\bf y}$). Upper case bold letters refer to matrices (e.g. ${\bf C}$). Unless otherwise stated, all parameters are complex numbers. The matrix inverse, transpose, Hermitian transpose, and conjugation are referred to as $(.)^{-1}$, $(.)^{T}$, $(.)^{H}$, $(.)^{\star}$, respectively. The matrix Kronecker product is given by $\otimes$. The statistical expectation operator is given as $E\{.\}$. The vectorized representation of a matrix is given by ${\rm vec}(.)$. The diagonal matrix consisting of only the diagonal entries of a square matrix is given by ${\rm diag}(.)$. The identity matrix is given by ${\bf I}$. The Kronecker delta function is given by $\delta_{ij}$. Real and complex numbers are represented as ${\mathbb R}$ and ${\mathbb C}$, respectively. Estimated parameters are denoted by a hat, $\widehat{(.)}$. All logarithms are to the base $e$.
\section{Data Model}
We briefly describe the data model of the radio interferometer in this section.  For more information about radio interferometry, the reader is referred to \cite{TMS} and for the data model in particular, \cite{HBS},\cite{HBS1}. A more signal processing oriented description is given in \cite{Boonstra03},\cite{Jeffs06}.

We consider the radio frequency sky to be composed of discrete sources, far away from the earth such that the approaching radiation from each one of them appears to be plane waves.
 Let the plane wave from the $i$-th source be decomposed to two orthogonal polarization directions ${\bf u}_i=[u_{xi}\ u_{yi}]^T$. The interferometric array consists of $N$ receiving elements or stations. At the $p$-th station, this plane wave causes an induced voltage, which is dependent on the beam attenuation as well as the radio frequency receiver chain attenuation. Normally, each station has dual polarized feeds. So the induced voltages at the $x$ and $y$ feeds, $\tilde{\bf v}_{pi}=[v_{xpi}\ v_{ypi}]^T$  due to source $i$ are given as in (\ref{ind}).
\beq \label{ind}
\tilde{\bf v}_{pi}={\bf J}_{pi} {\bf u}_i
\eeq
In (\ref{ind}), the complex interaction of the approaching radiation with the station beam shape as well as the remaining signal path is represented by the $2$ by $2$ Jones matrix ${\bf J}_{pi}$. If there are $K$ such sources, the total signal will be a superposition of $K$ such signals as in (\ref{ind}). Moreover, the receiver noise ${\bmath \nu}=[\nu_x\ \nu_y]^T$ is also added to this signal.

The signal of the $p$-th station is correlated with the signals of the other $N-1$ receivers at the correlator. Before this is done, each signal is given a delay correction depending on the direction on the sky being observed and also depending on the absolute position of the receiver on the earth.

After correlation, the correlated signal of the $p$-th station and the $q$-th station (named as the {\em visibilities} \cite{HBS}), ${\bf V}_{pq}=E\{ {\bf v}_{p} {\bf v}_{q}^H \}$ is given by (\ref{vispq}).
\beq \label{vispq}
{\bf V}_{pq}= \sum_{i=1}^{K} {\bf J}_{pi}({\bmath \theta}) {\bf C}_{i} {\bf J}_{qi}^{H}({\bmath \theta}) + {\bf N},\ \ p,q\in[1,\ldots,N]
\eeq
This is the full matrix measurement equation, developed in \cite{HBS}. In (\ref{vispq}), ${\bf J}_{pi}({\bmath \theta})$ and ${\bf J}_{qi}({\bmath \theta})$ are the Jones matrices describing the electromagnetic and electronic interaction of the plane wave of source $i$ with stations $p$ and $q$, respectively.  The parameter vector ${\bmath \theta} \in {\mathbb C}^P$ describes the unknown instrument model. The $2$ by $2$ noise matrix is given as ${\bf N}$.  We can only arrive at (\ref{vispq}) because the radiation emitted by the sources in the sky are uncorrelated. The {\em coherency} \cite{HBS}, ${\bf C}_i$  describes the intrinsic polarized radiation of the $i$-th source.

The instrumental properties (such as the beamshape, low noise amplifier gain, system frequency response etc.) and the path properties (such as tropospheric and ionospheric distortion etc.) are described by the Jones matrices ${\bf J}_{pi}({\bmath \theta})$ and ${\bf J}_{qi}({\bmath \theta})$  in (\ref{vispq}). Calibration is essentially finding the parameters ${\bmath \theta}$ ($P$ complex valued parameters or $2P$ real values parameters). There are numerous ways to parametrize these unknowns using the parameter set ${\bmath \theta}$. For a more specialized treatment of the parametrization as well as factoring the Jones matrices, the reader is referred to \cite{ND}.

To a lesser extent, the source information ${\bf C}_i$ is also unknown. However, at the initial stage, we can use prior information about the source or sky properties obtained by previous observations. 

Note that in (\ref{vispq}), the noise matrix  ${\bf N}={\bf 0}$ for $p\ne q$ if the noise at each receiver is uncorrelated. However in practice this does not hold for the following reasons:
\begin{itemize}
\item The integration time at the correlator has to be finite in order not to decorrelate the signal from the sources (or the sky). Hence there is always some receiver noise appearing even in the cross correlations, $p\ne q$.
\item The assumption that the sky is composed of a set of discrete sources is valid only up to a certain intensity level. There is low level diffused radiation from the sky, which appear especially in short baseline visibilities.
\item We select only $K$ brightest sources in our data model. However, the multitude of fainter sources that are ignored in (\ref{vispq}) contribute to the noise.
\end{itemize}

The vectorized form of (\ref{vispq}), ${\bf v}_{pq}=vec({\bf V}_{pq})$  can be written as in (\ref{vecvispq}) where ${\bf n}_{pq}=vec({\bf N})$.
\beq \label{vecvispq}
{\bf v}_{pq}=\sum_{i=1}^K {\bf J}_{qi}^{\star}({\bmath \theta}) \otimes {\bf J}_{pi}({\bmath \theta}) vec({\bf C}_{i}) + {\bf n}_{pq}
\eeq

Ignoring the autocorrelations where $p=q$, stacking up all cross correlations as ${\bf y}=[{\bf v}^T_{12}\ {\bf v}^T_{13}\ldots \ldots {\bf v}^T_{(N-1)N}]^T$, ${\bf y} \in {\mathbb C}^M$, we get (\ref{obs}).
\beq \label{obs}
{\bf y}=\sum_{i=1}^K {\bf s}_i({\bmath \theta}) + {\bf n}
\eeq

The size of ${\bf y}$, $M$, in (\ref{obs}) is at most $2N(N-1)$ provided all cross correlations are used. Typically, the number of parameters in ${\bmath \theta}$, $P$, is proportional to $KN$. So for large enough $N$ and small enough $K$, we have enough constraints to estimate ${\bmath \theta}$. The non-linear functions ${\bf s}_i({\bmath \theta})$ correspond to the contribution of each source to the observation. In previous formulations of the same problem, \cite{Jeffs06},\cite{Tol05}, the noise ${\bf n}$ has been ignored because only the cross correlations are used. However, we stress that in our formulation, we consider ${\bf n}$ to be a Gaussian random variable with zero mean and covariance  ${\bmath \Pi}$ ($M\times M$ matrix), i.e., ${\bf n} \sim {\mathcal N}({\bf 0},{\bmath \Pi)}$.

Note that (\ref{obs}) is a superposition of $K$ non linear signals, with unknown parameters, which is exactly the problem considered in \cite{Fed88}. The present calibration schemes estimate $\bmath \theta$ as the least squared error estimate, typically using a gradient based optimization algorithm like LM algorithm. 
\beq \label{mltheta}
\widehat{\bmath \theta}=\underset{\bmath \theta}{\rm arg\ min}\|{\bf y}-\sum_{i=1}^K {\bf s}_i({\bmath \theta})\|^2
\eeq
If the cost function is ${\bmath \phi}({\bmath \theta})=\|{\bf y}-\sum_{i=1}^K {\bf s}_i({\bmath \theta})\|^2$, at the $k$-th iteration, we estimate
\beq \label{lm0}
{\bmath \theta}^{k+1} ={\bmath \theta}^{k} - ({\bmath \nabla}_{\bmath \theta} {\bmath \nabla}^T_{\bmath \theta} {\bmath \phi}({\bmath \theta}) +\lambda {\bf H})^{-1} {\bmath \nabla}_{\bmath \theta} {\bmath \phi}({\bmath \theta})|_{{\bmath \theta}^k}
\eeq

In (\ref{lm0}), ${\bmath \nabla}_{\bmath \theta}$ is the gradient with respect to ${\bmath \theta}$ and $\lambda$ is a regularization parameter. The matrix ${\bf H}={\rm diag}({\bmath \nabla}_{\bmath \theta} {\bmath \nabla}^T_{\bmath \theta} {\bmath \phi}({\bmath \theta}))$ is the diagonal of the Hessian matrix. Given suitable initial values, (\ref{lm0}) should converge to the global optimum.
However, (\ref{lm0}) suffers from the same set of problems faced with any non linear optimization problem, i.e., convergence to local minima, slow convergence and heavy computational cost. In the next section we shall investigate the application of the EM algorithm to overcome some of these problems.
\section{The EM and SAGE Algorithms \label{emsage}}
In this section, we first proceed to apply the EM algorithm to (\ref{obs}), in a similar way as done in \cite{Fed88}. We do this before applying the SAGE algorithm to clarify the presentation. 
\subsection{EM Algorithm}
The key step in applying the EM algorithm is to define a complete data set ${\bf x}$ from the observed data ${\bf y}$. The obvious choice would be to associate each source with a complete data space $\tilde{\bf x}=[\tilde{\bf x}^T_1\ \tilde{\bf x}^t_2\ldots \tilde{\bf x}^T_K]^T$, with each component as in (\ref{cdata1}), such that ${\bf y}=\sum_{i=1}^{K}\tilde{\bf x}_i$.
\beq \label{cdata1}
\tilde{\bf x}_i={\bf s}_i({\bmath \theta}_i)+\tilde{\bf n}_i
\eeq

Note that in (\ref{cdata1}), we have  assumed the contribution of the $i$-th source depends only on a subset of parameters ${\bmath \theta}_i$, not the full set of parameters ${\bmath \theta}$. In other words, we partition the parameter space to $K$ components as ${\bmath \theta}=[{\bmath \theta}^T_1\  {\bmath \theta}^T_2\ldots {\bmath \theta}^T_K]^T$. This is justified because each source is at a unique direction on the sky. Even though the signal path for a given station is common for all sources, the different directions and the rotation of the sky makes this assumption justifiable. The noise contribution $\tilde{\bf n}_i$ is such that the total noise is decomposed into $K$ noise sources.
\beq \label{cnoise1}
{\bf n}=\sum_{i=1}^K \tilde{\bf n}_i,\ \ E\{\tilde{\bf n}_i\tilde{\bf n}^H_j\}= \beta_i \delta_{ij} {\bmath \Pi},\ \ \sum_{i=1}^K \beta_i=1
\eeq

The $\beta_i$s form an affine combination and we are free to choose them. Typically, we can associate stronger sources with lower noise, hence low $\beta_i$.
Given the complete data ${\bf x}$, we get the observed data as in (\ref{compo}) where ${\bf G}$ is a block matrix with $K$ identity matrices.
\beq \label{compo}
{\bf y}=[{\bf I}\ {\bf I}\ \ldots {\bf I}] {\bf x}={\bf G}{\bf x}
\eeq

Having this setup, it is rather straightforward to apply the EM algorithm to our problem as in \cite{Fed88}.

{\em E Step}: We find the conditional mean of $\widehat{\tilde{\bf x}}_i=E\{\tilde{\bf x}_i| {\bf y}, {\bmath \theta}^k\}$. Taking into account that ${\bf y}$ and ${\bf x}$ are jointly Gaussian, we get
\beq \label{cmx}
\widehat{\tilde{\bf x}_i}= {\bf s}_i({\bmath \theta}^k_i) + \beta_i ( {\bf y}-\sum_{l=1}^K {\bf s}_l({\bmath \theta}_l^k))
\eeq

{\em M Step}: For the $k+1$-th iteration, we find ${\bmath \theta}_i^{k+1}$ that minimizes the cost ${\bmath \phi}_i({\bmath \theta}_i^{k+1})=\|\widehat{\tilde{\bf x}_i}-{\bf s}_i({\bmath \theta}_i^{k+1})\|^2$, given by:
\beq \label{mstep}
{\bmath \theta}_i^{k+1}={\bmath \theta}_i^k - ({\bmath \nabla}_{{\bmath \theta}_i} {\bmath \nabla}^T_{{\bmath \theta}_i} {\bmath \phi}_i({\bmath \theta}_i) +\lambda {\bf H}_i)^{-1} {\bmath \nabla}_{{\bmath \theta}_i} {\bmath \phi}_i({\bmath \theta}_i)|_{{\bmath \theta}^k_i}
\eeq
where ${\bf H}_i={\rm diag}({\bmath \nabla}_{{\bmath \theta}_i} {\bmath \nabla}^T_{{\bmath \theta}_i} {\bmath \phi}_i({\bmath \theta}_i))$.
We repeat the above two steps starting from iteration $k=1$ until convergence or an upper limit has reached. At each iteration, we update each source, so $i$ goes from $1$ to $K$.

\subsection{SAGE Algorithm}
Next, we investigate the application of the SAGE algorithm to our problem. As before we need to find a complete data space (or a hidden data space as defined in \cite{Fess94}). Similar to \cite{Fess94}, we select the hidden data space ${\bf x}^S$ as in (\ref{hid}).
\beq \label{hid}
{\bf x}^S= {\bf s}_i({\bmath \theta}_i) + {\bf n}
\eeq
This gives the observed data ${\bf y}$ as in (\ref{sageo}).
\beq \label{sageo}
{\bf y}= {\bf x}^S + \sum_{l=1,l\ne i}^{K}  {\bf s}_l({\bmath \theta}_l) 
\eeq
Note that in (\ref{hid}) and (\ref{sageo}), we have selected the index set \cite{Fess94}, $S$ to be the $i$-th source. Moreover, we have associated all the noise to ${\bf x}^S$, unlike in the classic EM algorithm. Once again, we arrive at the following EM scheme:

{\em SAGE E Step}: We find the conditional mean of ${\bf x}^S=E\{ {\bf x}^S| {\bf y}, {\bmath \theta}^k\}$.
\beq \label{sagecmx}
\widehat{{\bf x}^S}= {\bf s}_i({\bmath \theta}^k_i) + ( {\bf y}-\sum_{l=1}^K {\bf s}_l({\bmath \theta}_l^k)) = {\bf y}-\sum_{l=1,l\ne i}^K {\bf s}_l({\bmath \theta}_l^k)
\eeq

{\em SAGE M Step}: For the $k+1$-th iteration, ${\bmath \theta}_i^{k+1}$ that minimizes the cost ${\bmath \phi}_S({\bmath \theta}_i^{k+1})=\|\widehat{{\bf x}^S}-{\bf s}_i({\bmath \theta}_i^{k+1})\|^2$. This is similar to (\ref{mstep}). As before, we iterate from $k=1$ to an upper limit. At each iteration, we change the index set $S$ to update all or some sources. 

Note that instead of partitioning per source, we could also perform the partitioning to include more than one source. This would be better if some sources are closer in the sky and hence share some parameters.
\subsection{Computational Cost\label{comp_cost}}
The computational cost of direct estimation using (\ref{lm0}) and the EM algorithmic approach can be compared as follows. Solving (\ref{lm0}) (without calculating the inverse) involves the solution of a linear system of order $KN$. So the computational cost of the direct approach is ${\mathcal O}((KN)^2)$. On the other hand, the computational cost of solving (\ref{mstep}), $K$ times, is $K {\mathcal O}(N^2)$. Thus, we gain a factor $K$ by using the EM algorithm. Furthermore, we can increase this gain if the convergence of the EM approach is faster (fewer iterations). 

\subsection{Comparison With "Peeling"\label{peeling}}
As described in \cite{Jeffs06} in detail, {\em Peeling} is the conversion of the $K$ source model in (\ref{obs}) to a series of single source calibration problems, exploiting the temporal diversity due to the rotation of the earth. The steps taken in {\em Peeling} can be briefly given as follows:
\begin{itemize}
\item Out of sources, $i\in[1,K]$, select the strongest source (say $i=q$).
\item Optionally, subtract the contributions of the remaining sources from (\ref{obs}), using an approximate a priori instrument model.
\item Multiply (\ref{obs}) by a diagonal matrix ${\bf F}$ such that the $q$-th source is at the phase center. The matrix ${\bf F}$ is computed for given time and the absolute position of the $q$-th source in the sky.
\item Provided that the remaining sources are weak enough, over a finite time interval, the contribution of those sources in (\ref{obs}) are averaged out, while the contribution of the $q$-th source remains constant.
\item Ignore the contribution from the other sources $i\in[1,K],i\ne q$ in (\ref{obs}) and solve for the parameters of the $q$-th source. Subtract this from (\ref{obs}). Now, we have $K-1$ sources left and we repeat the whole procedure.
\end{itemize}

As seen from above, the application of the proposed algorithm does not rely on the weaker sources being averaged out. Moreover, the proposed algorithm does not require an a priori instrument model. When we have equally strong sources, {\em Peeling} might not work satisfactorily compared to the proposed algorithm.

\section{Statistical Model Order Selection\label{stat}}
So far in our analysis, we have assumed the number of sources, $K$, in  (\ref{vispq}), is known a priori. To some extent, this is true, given prior observational data and receiver noise characteristics. It is straightforward to find $K$ sources that has sufficient SNR given the aforementioned information. However, in situations where the receiver antenna beamshape varies with time due to earth rotation (as in LOFAR), this is hard to predict (e.g. some sources might go close to, or, even below, the horizon). In this situation without a priori knowledge, we could use information theoretic criteria to find the optimal $K$ for a given observation. In this section, we describe the use of Akaike's Information Criterion (AIC) \cite{AIC} for this purpose. There are also alternative criteria, see for instance \cite{Wax85} for more information.

From (\ref{obs}), the likelihood of ${\bf y}$ is given by (\ref{lkhood}).
\beq \label{lkhood}
f({\bf y}|{\bmath \theta})=\frac{1}{\pi^M |{\bmath \Pi}|} \exp(-({\bf y}-\sum_{i=1}^K {\bf s}_i({\bmath \theta}))^{H} {\bmath \Pi}^{-1} ({\bf y}-\sum_{i=1}^K {\bf s}_i({\bmath \theta})))
\eeq

Assuming the noise to be white, ${\bmath \Pi}=\sigma^2{\bf I}$, we get the simplified log-likelihood $L({\bmath \theta})$ as in (\ref{llh}).
\beqn \label{llh}
\lefteqn {L({\bmath \theta})=\log f({\bf y}|{\bmath \theta})}\\\nonumber
&&=-M \log \pi -M\log \sigma^2\\\nonumber
&&-\frac{1}{\sigma^2}({\bf y}-\sum_{i=1}^K {\bf s}_i({\bmath \theta}))^{H}({\bf y}-\sum_{i=1}^K {\bf s}_i({\bmath \theta})) 
\eeqn

The maximum likelihood estimate for the noise variance $\sigma^2$ (given ${\bmath \theta}$) is given by (\ref{mlsig}).
\beq \label{mlsig}
\widehat{\sigma^2}=\frac{1}{M}({\bf y}-\sum_{i=1}^K {\bf s}_i(\widehat{\bmath \theta}))^{H}({\bf y}-\sum_{i=1}^K {\bf s}_i(\widehat{\bmath \theta}))
\eeq

Using (\ref{mlsig}) in (\ref{llh}), we arrive at (\ref{mlllh}).
\beqn \label{mlllh}
\lefteqn {L(\widehat{\bmath \theta})=-M \log \pi -M}\\\nonumber
&&-M\log \bigl(\frac{1}{M}({\bf y}-\sum_{i=1}^K {\bf s}_i(\widehat{\bmath \theta}))^{H}({\bf y}-\sum_{i=1}^K {\bf s}_i(\widehat{\bmath \theta}))\bigr) 
\eeqn

Using (\ref{mlllh}), we get the Akaike's Information Criterion as (\ref{aic}). We select $K$ that gives the minimum value for (\ref{aic}).
\beq \label{aic}
AIC(K)=-2 L(\widehat{\bmath \theta}) + 2 (2P)
\eeq

\section{Numerical Example}
\begin{figure}[htbp]
\begin{minipage}[b]{0.98\linewidth}
\centering
\centerline{\epsfig{figure=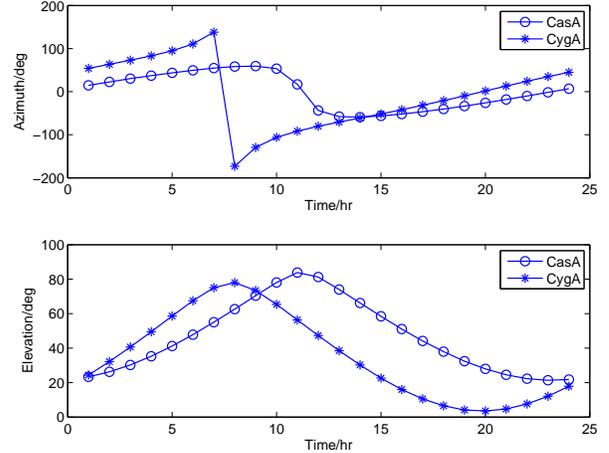,width=7.9cm}}
\end{minipage}
\caption{The positions of CasA and CygA on the sky in azimuth and elevation, for a geographic latitude of $53^{o}$.} \label{azel}
\end{figure}

We consider the calibration of some data obtained by the LOFAR test core station (CS1). This has $N=16$ dipoles (with dual polarization) acting as an interferometric array. Since each station is a single dipole, there is no beamforming and thus the whole sky (hemisphere) is observed. In this setting, the two brightest sources are Cassiopeia A (CasA) and Cygnus A (CygA), with intensities about 20000 Jy each at 50 MHz. The observation lasts for 24 hours. The correlator integration time is 30 sec. During the observation, the positions of the sources (azimuth and elevation) vary as shown on Fig. (\ref{azel}).

Since both CasA and CygA are equally bright, traditional algorithms such as peeling \cite{Jeffs06} will not work satisfactorily, as described in section \ref{peeling}. So we have a model with $K=2$ in (\ref{obs}). Note that as seen on Fig. (\ref{azel}), CygA goes very close to the horizon at one point. Around this time, the contribution from CygA is almost negligible due to the attenuation by the dipole beam. So, instead of using $K=2$, we should be using $K=1$. However, in this example, we only consider the data where both CasA and CygA are high in elevation (about 18 hours). Future work will address using e.g., (\ref{aic}) to determine this. 

The parametrization of the Jones matrices are done as follows: We consider each entry of the $2$ by $2$ matrix to be a parameter. For $K=2$ and $N=16$, there are $2\times16\times4=128$ parameters.
\beqn \nonumber \label{Jones}
{\bf J}_p({\bmath \theta}_i)=\left[ \begin{array}{cc}
J_{p11i} & J_{p12i}\\
J_{p21i} & J_{p22i}
\end{array} \right],\\
 {\bmath \theta}_i=[{\rm vec}({\bf J}_p({\bmath \theta}_i))^T,\ldots,]^T\ \forall p \in [1\ldots,N] 
\eeqn
The Jones matrices are initialized such that the diagonal entries each have a (real) value $0.0001$ and the off diagonal entries to be zero.

We consider the estimation of the parameters using (\ref{mltheta}) (Normal Algorithm) and using the SAGE algorithm. For the normal algorithm, we use $12$ and $24$ LM iterations to estimate $128$ parameters using (\ref{mltheta}). For the SAGE algorithm, we alternate between estimating parameters for CasA and CygA. In each iteration, we use $3$ LM iterations for the M step (\ref{mstep}). We use $4$ EM iterations, keeping the number of LM iterations at $12$. Yet, the SAGE algorithm is computationally less expensive than the normal algorithm with $12$ iterations as noted in section \ref{comp_cost}.

\begin{figure}[htbp]
\begin{minipage}[b]{0.98\linewidth}
\begin{minipage}[b]{0.48\linewidth}
\centering \centerline{\epsfig{figure=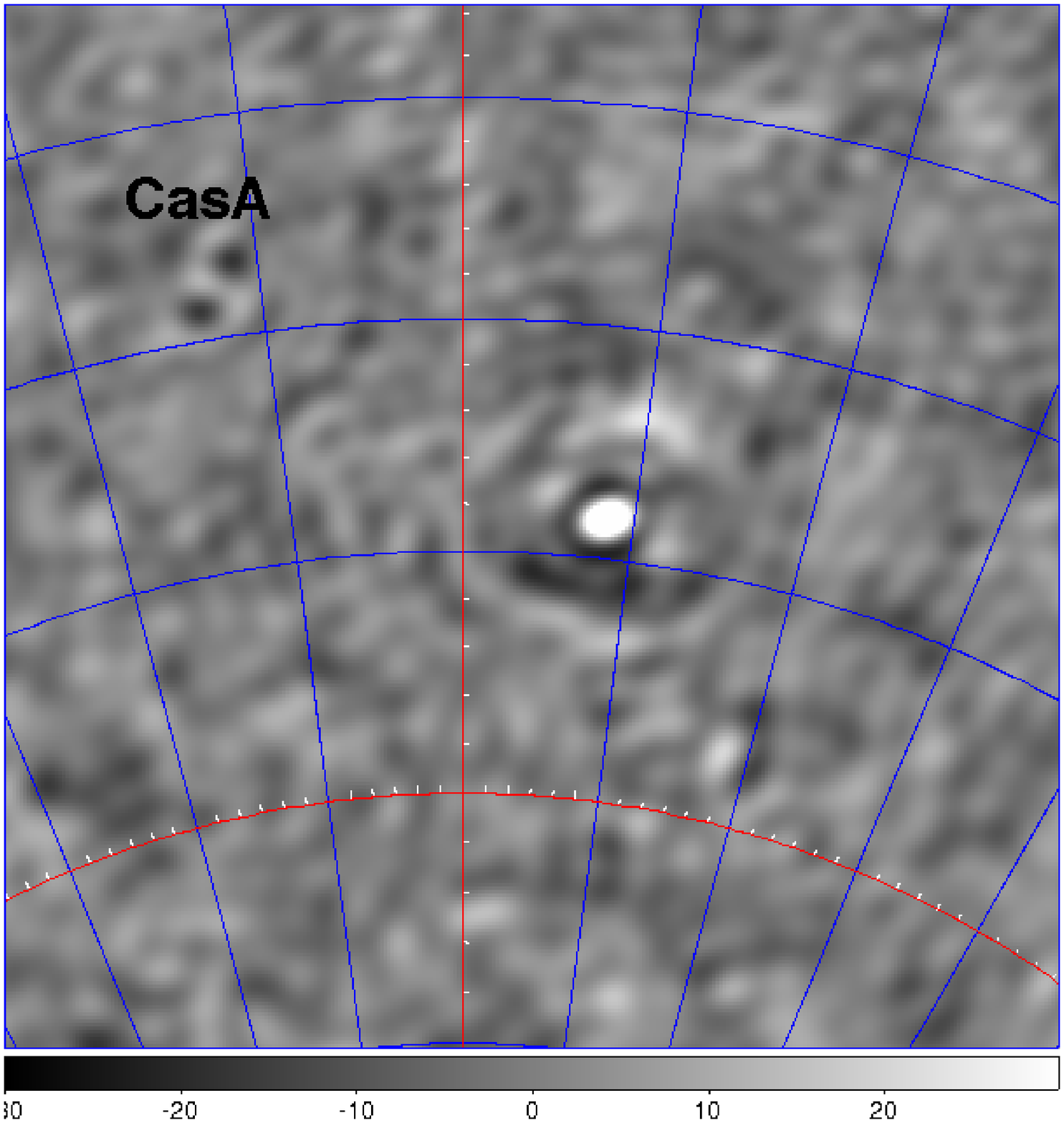,width=4cm}}
\vspace{0.2cm}\centerline{CasA}
\end{minipage}
\begin{minipage}[b]{0.48\linewidth}
\centering \centerline{\epsfig{figure=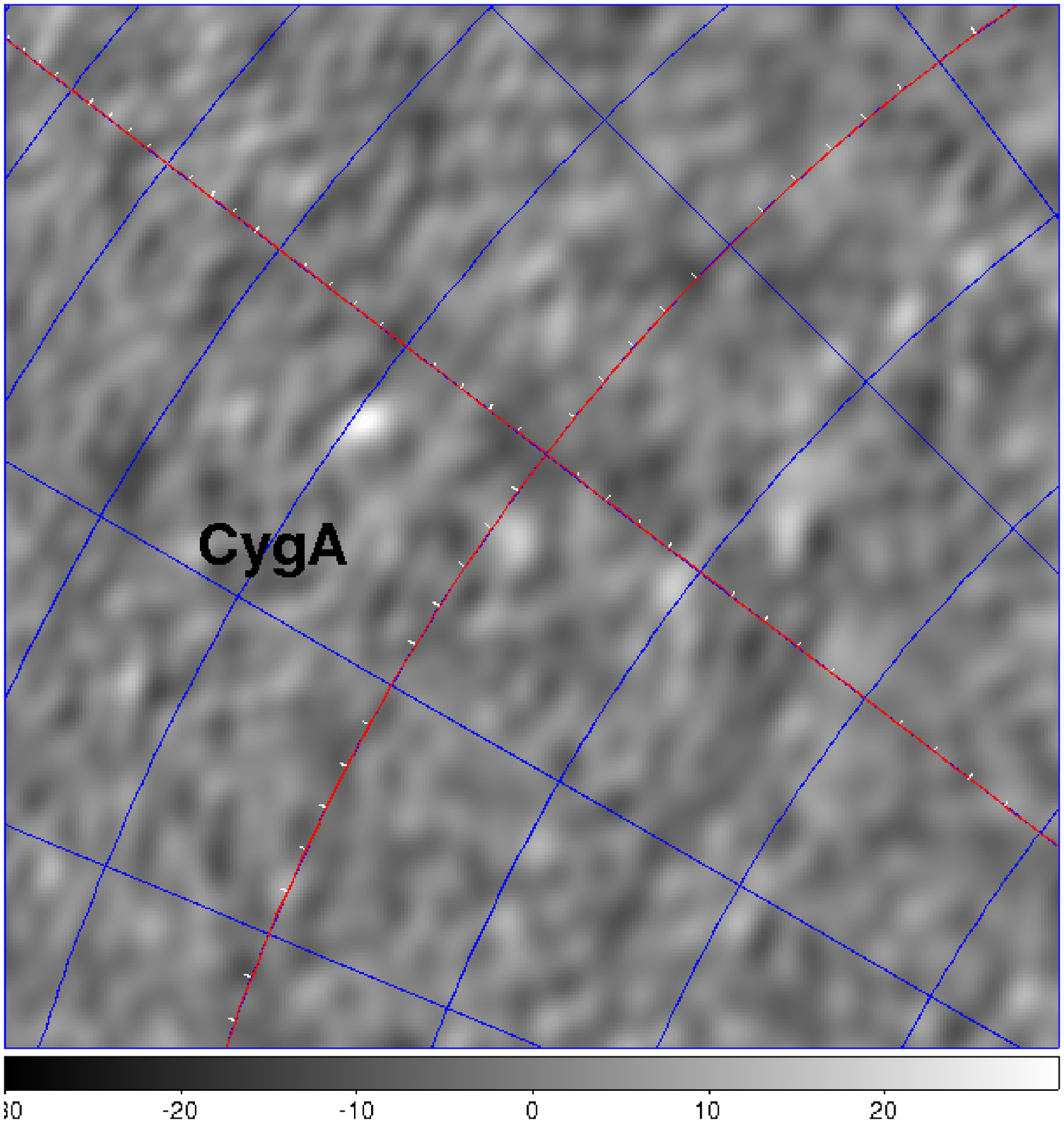,width=4cm}}
\vspace{0.2cm}\centerline{CygA}
\end{minipage}
\centerline{SAGE, 12 iterations}
\vspace{2pt}\\
\begin{minipage}[b]{0.48\linewidth}
\centering \centerline{\epsfig{figure=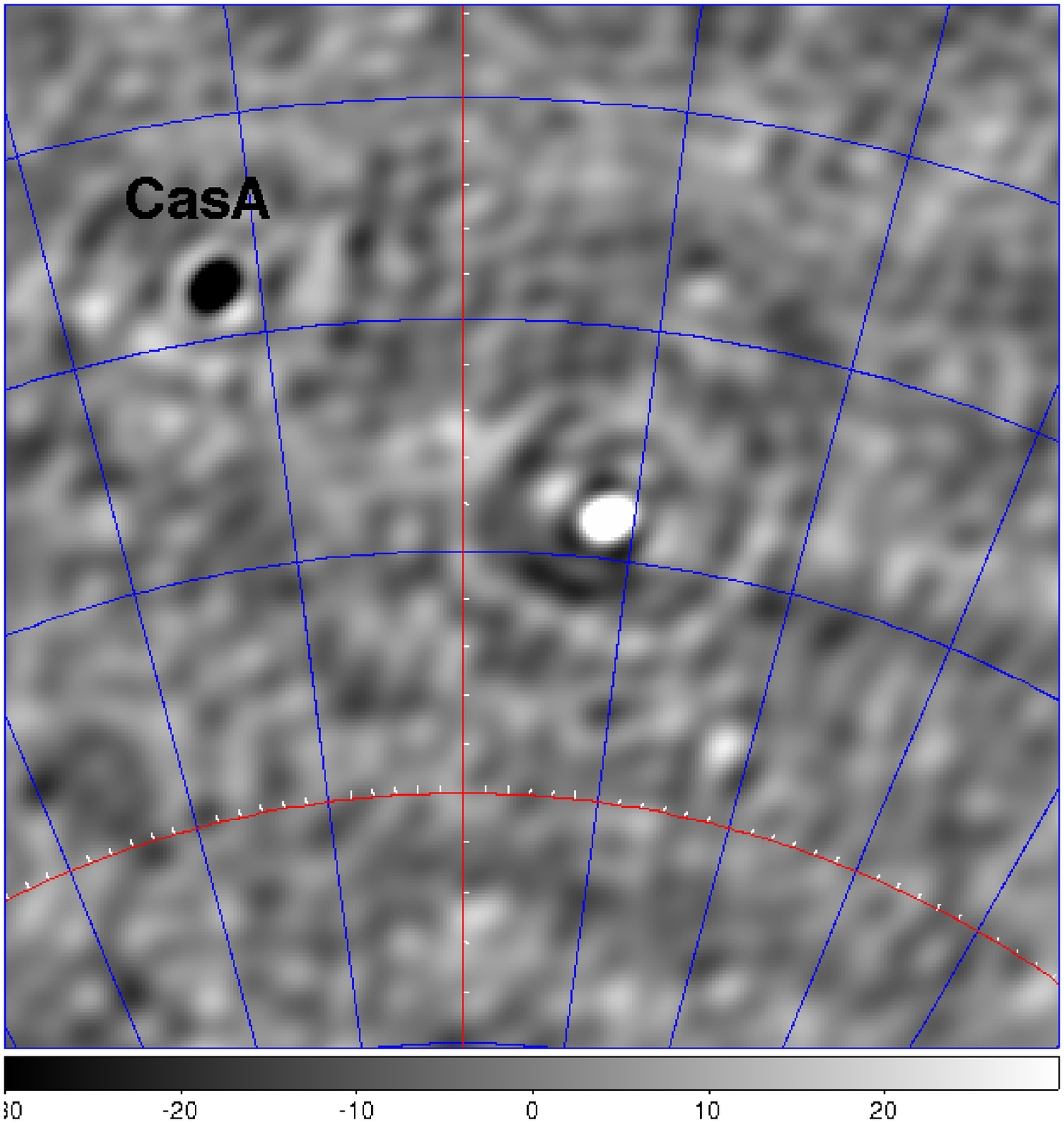,width=4cm}}
\vspace{0.2cm}\centerline{CasA}
\end{minipage}
\begin{minipage}[b]{0.48\linewidth}
\centering \centerline{\epsfig{figure=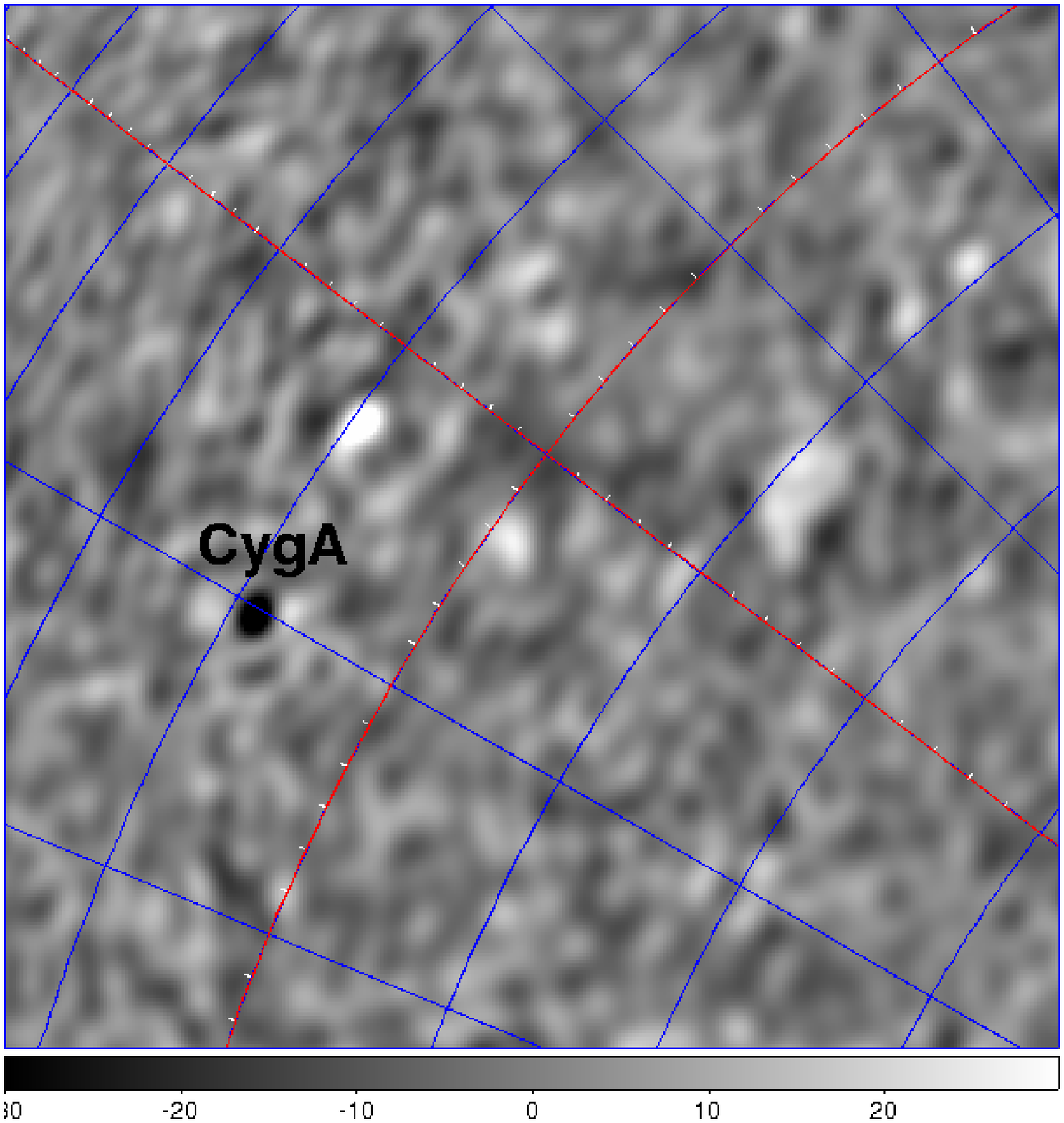,width=4cm}}
\vspace{0.2cm}\centerline{CygA}
\end{minipage}
\centerline{Normal, 12 iterations}
\vspace{2pt}\\
\begin{minipage}[b]{0.48\linewidth}
\centering \centerline{\epsfig{figure=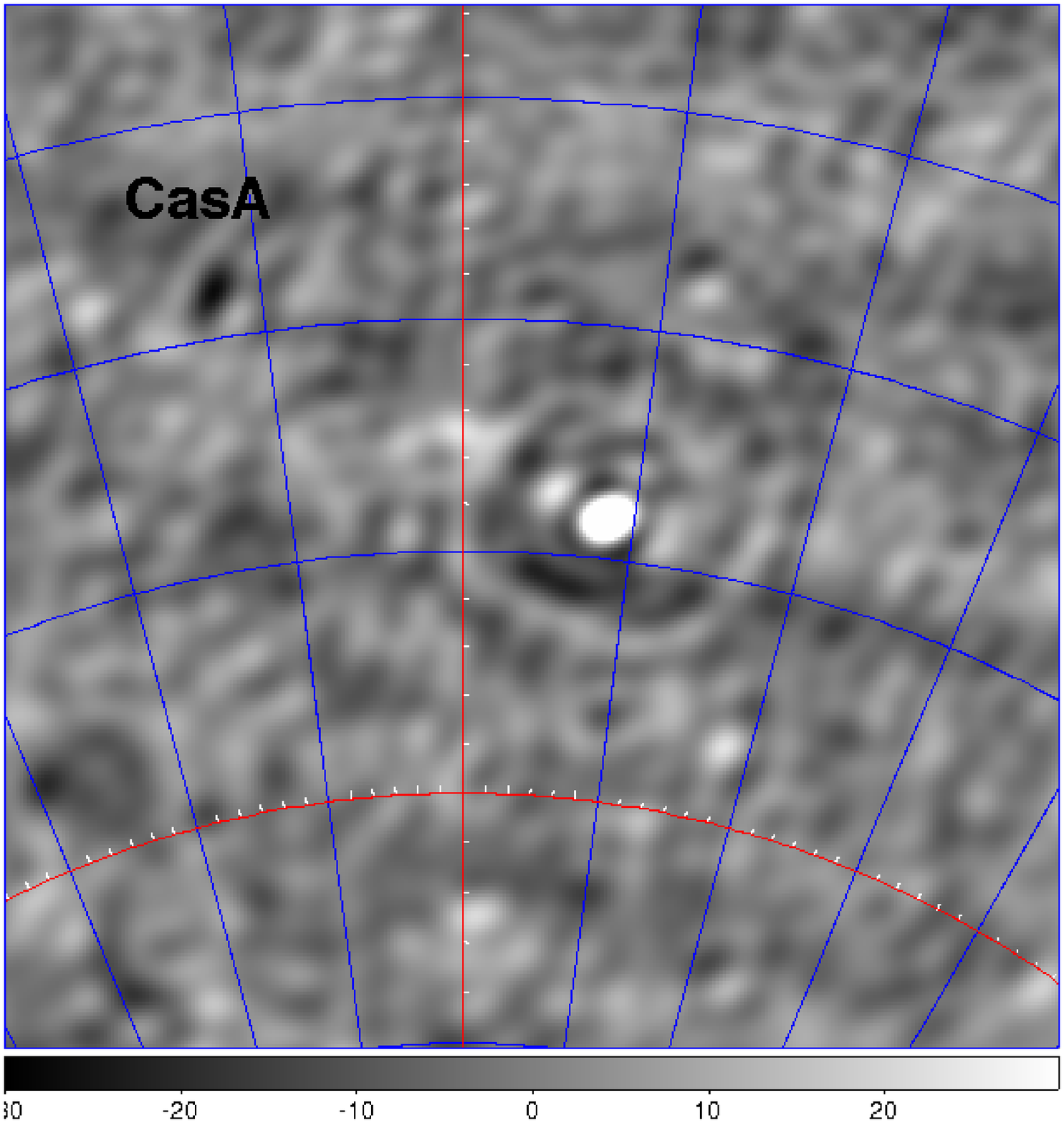,width=4cm}}
\vspace{0.2cm}\centerline{CasA}
\end{minipage}
\begin{minipage}[b]{0.48\linewidth}
\centering \centerline{\epsfig{figure=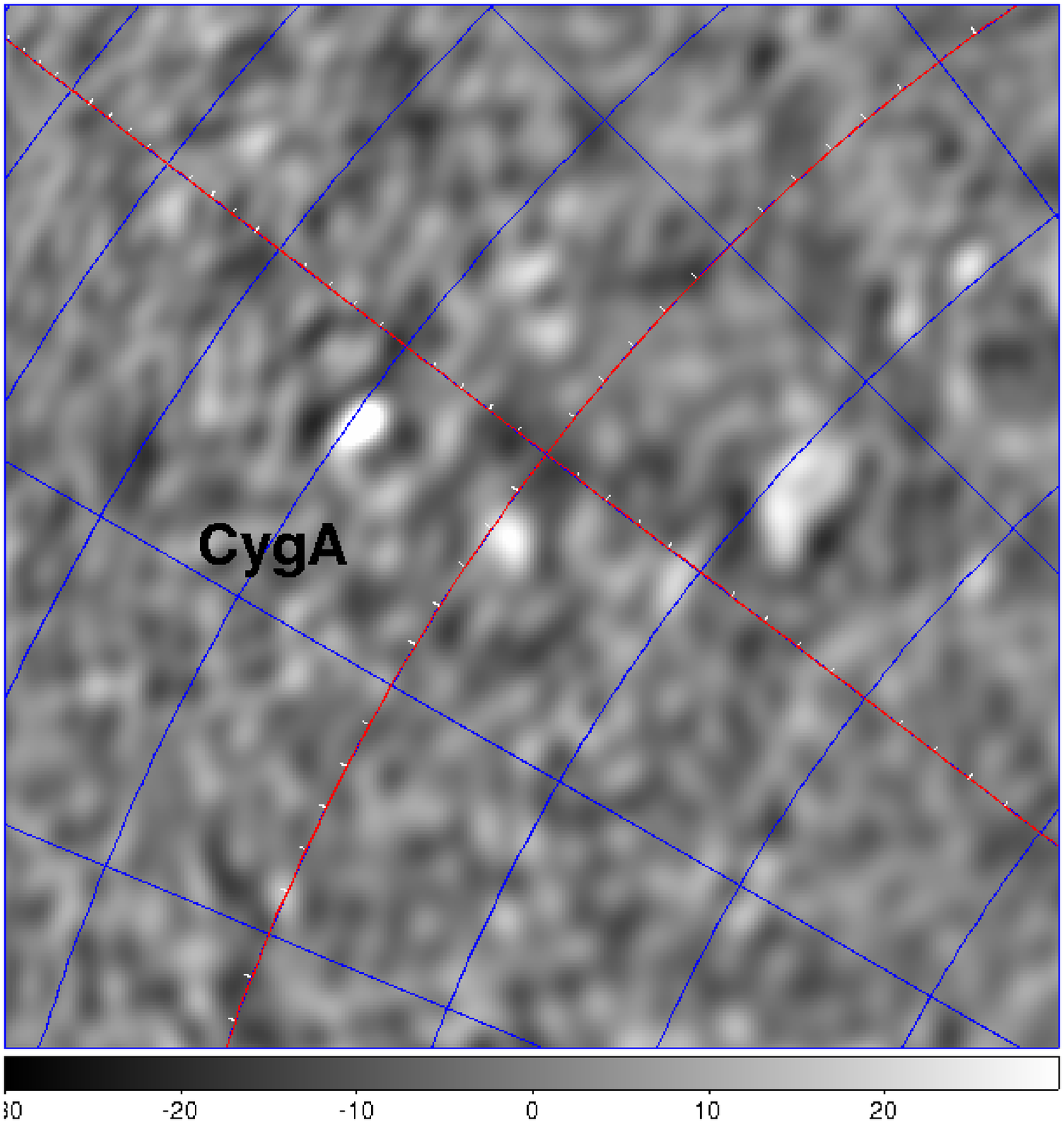,width=4cm}}
\vspace{0.2cm}\centerline{CygA}
\end{minipage}
\centerline{Normal, 24 iterations}
\end{minipage}
\caption{Images around CasA (left column) and CygA (right column) after applying the SAGE algorithm (first row), Normal algorithm, 12 iterations (second row) and Normal algorithm, 24 iterations (bottom row). The residual of CasA is seen at top left on the images in the left column. The residual of CygA is seen at center left on the images in the right column. The grid lines correspond to sky coordinates: right ascension and declination.}
\label{figcasa}
\end{figure}

Once we have estimated ${\bmath \theta}$, we make images of the residual, i.e. ${\bf y}- \sum_{i=1}^K {\bf s}_i(\widehat{\bmath \theta})$. We also correct the residual using the estimated ${\bmath \theta}$ \cite{SBY}. We have given the images made by the normal algorithm and the SAGE algorithm on Fig. \ref{figcasa}. These images show an area around CasA and CygA, respectively. Perfect subtraction should leave no residual from both these sources. However, we see that there is about 1\% (of the original value) peak residual left by using the normal algorithm with $12$ iterations. On the other hand, the SAGE algorithm and the normal algorithm with $24$ iterations reduce this residual to 0.1\% level. Moreover, fainter, known sources can also be seen on both images. Closer scrutiny reveals that the remaining sources are fainter in the results obtained using the SAGE algorithm. This is due to over subtraction of the fluxes of the remaining sources and in fact, we could reduce the number of SAGE iterations, to overcome this effect. Future work will address determining the correct number of iterations to avoid over subtraction.

In order to have a quantitative handle on the results, we have also calculated the root mean square (rms) value of the residual on these images. For an image with $L_1\times L_2$ pixels, the rms value, $\eta$ can be defined as in (\ref{rms}). In (\ref{rms}), the value at pixel $i,j$ is given by $z_{ij}$.
\beq \label{rms}
\eta=\sqrt{\frac{1}{L_1 L_2} \sum_{i=1}^{L_1}\sum_{j=1}^{L_2} z_{ij}^2 }
\eeq

\begin{figure}[htbp]
\begin{minipage}[b]{0.98\linewidth}
\centering
\centerline{\epsfig{figure=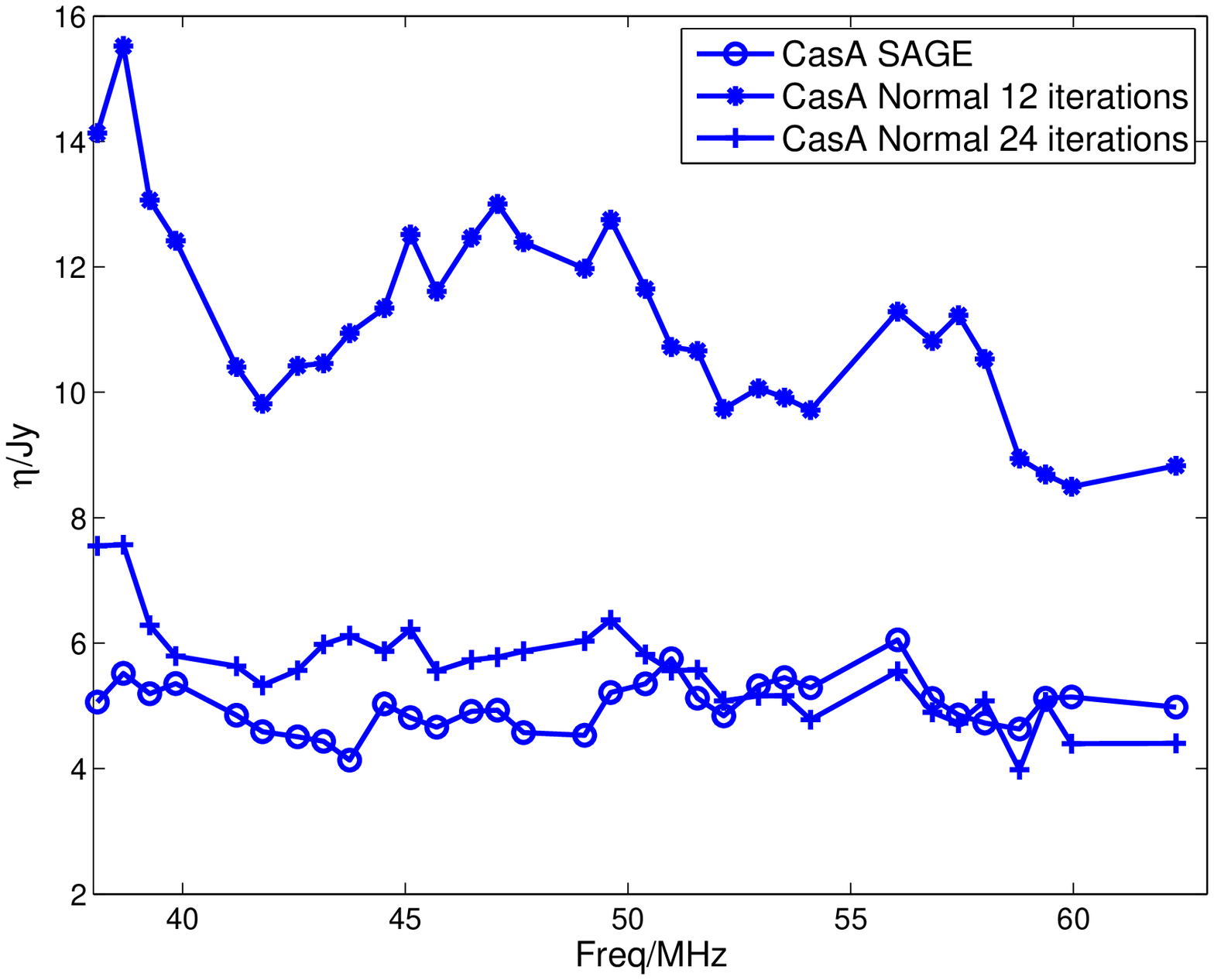,width=7.9cm}}
\vspace{0.5cm}\centerline{CasA}\smallskip
\centering
\centerline{\epsfig{figure=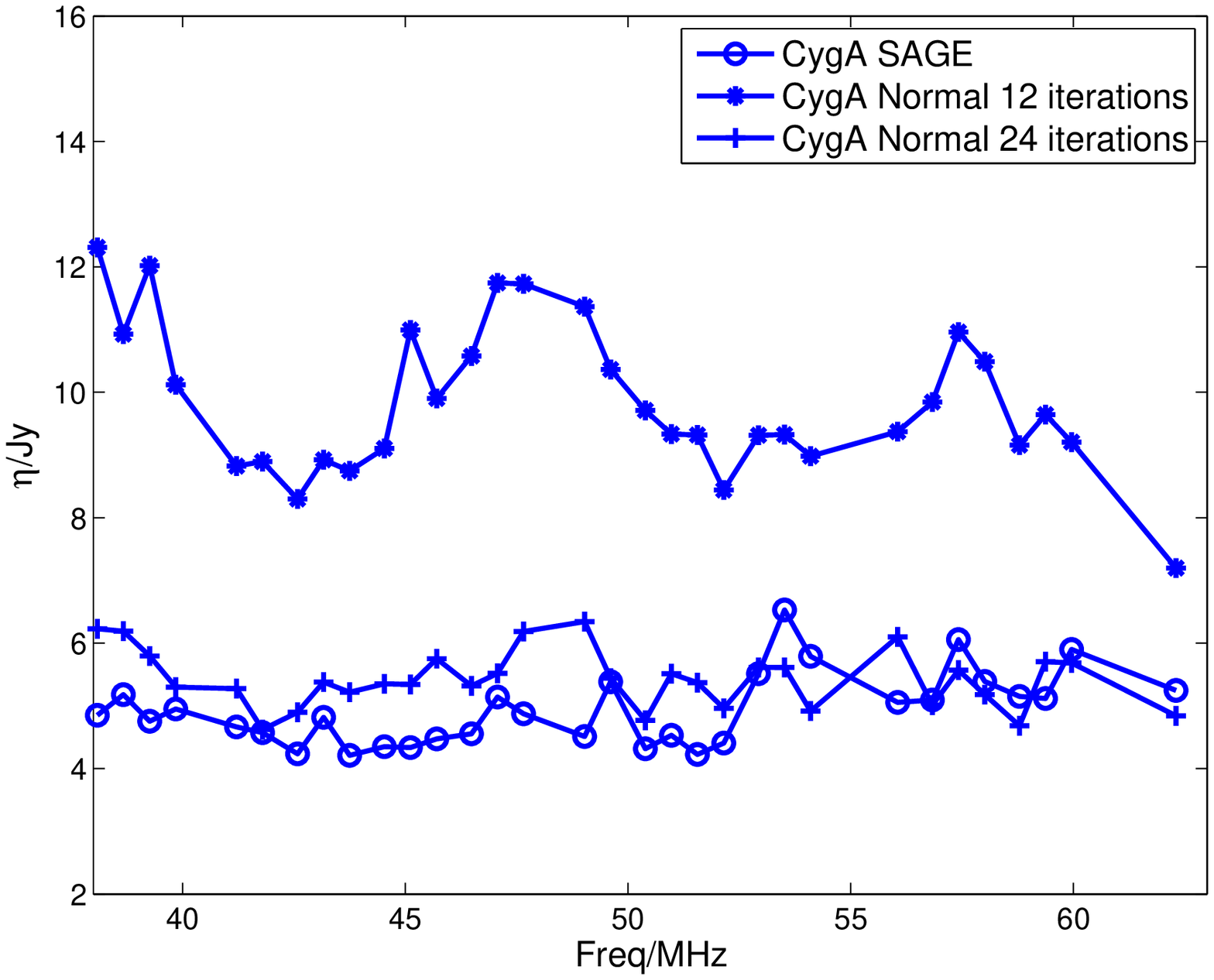,width=7.9cm}}
\vspace{0.5cm}\centerline{CygA}\smallskip
\end{minipage}
\caption{Pixel rms values around CasA and CygA using the normal calibration algorithm and the SAGE calibration algorithm. For equal number of iterations, the normal algorithm has higher residual compared to the SAGE algorithm. With higher number of iterations, the normal algorithm gives comparable performance.} \label{eta}
\end{figure}

We have evaluated (\ref{rms}) on images centered around CasA and CygA, with $64$ by $64$ pixels in size, for about $30$ different frequencies around 50 MHz.
The results can be seen on Fig. \ref{eta} for both CasA and CygA. It is clearly seen that for the same number of iterations, the SAGE algorithm performs better. The normal algorithm needs about more iterations to have the same performance as the SAGE algorithm.
\section{Conclusions}
We have presented the application of the SAGE algorithm for calibration of radio interferometric arrays. This is an improvement over the generally used direct optimization methods in performance as well as in computing cost, as seen from the results. We have only given some initial results of using the SAGE algorithm on real data. One of the fundamental assumptions made in this paper was that the noise is white and Gaussian.  Future work will address exact characterization of the noise and adaptation of the SAGE algorithm especially when the noise is non Gaussian. Moreover, future work will address situations where we have more than $2$ strong sources.

\section{Acknowledgment}
The first author would like to thank ASTRON for kind hospitality and NOVA for support. This work 
was also supported by LOFAR and SNN. LOFAR is being funded by the European Union, European Regional Development Fund, and by ``Samenwerkingsverband Noord-Nederland'', EZ/KOMPAS. We acknowledge Ronald Nijboer for reviewing an earlier version of this paper.

\nocite{*}
\bibliographystyle{IEEE}

\end{document}